\documentclass[12pt,english]{article}
\usepackage{mathpazo}
\usepackage[T1]{fontenc}
\usepackage[latin9]{inputenc}
\usepackage{geometry}
\usepackage{fancyhdr}
\usepackage{amsmath}
\usepackage{amssymb}
\usepackage{esint}

\makeatletter
\newcommand{\ve}[1]{\ensuremath{\mbox{\boldmath$#1$}}}

\usepackage{babel}
\makeatother

\usepackage{babel}
\begin{document}
\begin{flushright}
\thispagestyle{empty}Reeks,M.W. (2014) arXiv:1409.5714 {[}physics.flu-dyn{]}
\par\end{flushright}

\begin{center}
{\LARGE \vspace{6mm}
The concept of particle pressure in a suspension of particles in a
turbulent flow\vspace{3mm}
}
\par\end{center}{\LARGE \par}

\begin{center}
{\large Michael W Reeks }\\
{\large{} School of Mechanical \& Systems Engineering, }\\
{\large{} University of Newcastle, UK}\vspace{5mm}

\par\end{center}
\begin{abstract}
The Clausius Virial theorem of Classical Kinetic Theory is used to
evaluate the pressure of a suspension of small particles at equilibrium
in an isotropic homogeneous and stationary turbulent flow. It then
follows a similar approach to the way Einstein \cite{Einstein1905}
evaluated the diffusion coefficient of Brownian particles (leading
to the Stokes-Einstein relation) to similarly evaluate the long term
diffusion coefficient of the suspended particles. In contrast to Brownian
motion, the analogue of temperature in the equation of state which
relates pressure to particle density is not the kinetic energy per
unit particle mass except when the particle equation of motion approximates
to a Langevin Equation. 
\end{abstract}
In this short paper I reexamine how in Reeks (1991)\cite{Reeks91},
the Clausius Virial Theorem was used to obtain the equation of state
for a suspension of small particles at equilibrium in a statistically
stationary homogeneous isotropic turbulent flow. The idea of using
the Virial Theorem came from Fowler's classic book on Statistical
Physics \cite{Fowler66} where it was used to derive the equation
of state of a non ideal gas. There is an obvious analogy between molecules
in a gas and particles suspended in a turbulent gas flow. And indeed
in applying the Virial Theorem, it doesn\textquoteright{}t matter
that the forces on the individual particles are different from those
of the gas molecules or that the kinetic energy of the molecules is
derived from their collisions with one another and that for a dilute
suspension of particles, it results from their interaction with the
underlying turbulent carrier gas flow. In that respect the theorem
is completely general. Both systems are considered at equilibrium
($t\rightarrow\infty$) when particles / molecules are uniformly mixed
in terms of concentration and kinetic energy (temperature). As with
molecules in a gas, the suspended particles are confined within the
walls of some container that impose an external stress on the particles
that is equal and opposite to the pressure exerted by the suspended
particles. Because the particles are in equilibrium, the pressure
is the same everywhere internally and the same stresses that apply
at the walls as physical boundaries apply to any geometrical surface
internally (i.e within the container). %
\footnote{Of course dealing with internal \emph{geometrical} surfaces gets round
the problem that the physical boundaries influence the carrier flow.
We would naturally suppose that this has a negligible effect on the
particle. i.e. it is an extremely thin near wall boundary layer and
the particle inertia is so great that the particles are unaffected.
Alternatively we might consider a semi-impermeable wall that is permeable
to the carrier flow but impermeable to the suspended particles.%
}\\

We thus consider the motion of an individual particle in a suspension
of $N$ particles all of the same mass $m$ at equilibrium in a statistically
stationary homogeneous isotropic turbulent flow. This particle has
a velocity $\ve\upsilon$ and position $\ve x$ at time $t$ and is
subject to a resistive force (per unit particle mass) proportional
to its velocity,$-\beta\ve\upsilon$ where $\beta$ is a constant,
and a driving force (per unit mass) due to the turbulence $\ve f(t)$
measured along its trajectory at time $t$ which is fluctuating in
time on a time scale $\sim\tau_{f}$ with an average value of zero.
The equation of motion of motion of this particle is thus explicitly
\begin{equation}
\frac{d\ve\upsilon}{dt}=-\beta\ve\upsilon+\ve f(t)+m^{-1}\ve F_{e}\;;\;\frac{d\ve x}{dt}=\ve\upsilon\label{eq:particle equation of motion}
\end{equation}
where $\ve F_{e}$ is an external force acting on the individual particles
which is everywhere zero except at the walls where it is equal and
and opposite to the force imposed by the particles impacting at the
walls and the source of the particle pressure. For molecules in a
gas, $\ve F_{e}$ also accounts for the inter molecular forces and
is therefore non-zero internally. We assume here like an ideal gas,
there are no inter particle forces . $\beta^{-1}$ we refer to as
the particle  response time, measuring the response of the particle
to changes in the flow occurring on a timescale of $\tau_{f}$ . $(\beta\tau_{f})^{-1}$
is thus a measure of the particle inertia and is referred to as the
particle Stokes number $St$. $St\ll1$ corresponds to a particle
of weak inertia where the particle almost follows the carrier flow,
and $St\gg1$ defines a particle with a high inertia in which $\ve f(t)$
is effectively white noise, i.e on the timescale of the particle motion
$\beta^{-1}$. In the case of small particles with a low particle
Reynolds number $Re_{p}$, $\ve f(t)=\beta\ve u(t)$ where $\ve u$(t)
is the local carrier flow velocity (along its trajectory at time $t$)
so that the net force (per unit mass) due to the carrier flow on a
particle with velocity $\ve\upsilon$ at time $t$ is given by Stokes
drag $\beta(\ve u-\ve\upsilon)$. Thus Eq.(\ref{eq:particle equation of motion})
is meant to cover the entire range of Stokes numbers ($0\leqslant St\leqslant\infty)$.
In general $\beta$ is a function of the particle Reynolds number
$Re_{p}$ (see Reeks \cite{Reeks80} for the value of $\beta$ for
high inertia particles).\\
Multiplying Eq.(\ref{eq:particle equation of motion}) by $\frac{1}{2}x_{i}$,
rearranging using product differentiation, summing over $i$, and
rearranging the equation so that all the time derivative quantities
are on the left hand side, we have 
\begin{equation}
\frac{1}{4}\frac{d^{2}x^{2}(t)}{dt^{2}}+\frac{1}{4}\beta\frac{dx^{2}(t)}{dt}=\frac{1}{2}\upsilon^{2}+\frac{1}{2}\ve x(t)\cdot\ve f(t)+\frac{1}{2}m^{-1}\ve F_{e}\cdot\ve x(t)\label{eq:}
\end{equation}
where $x=\left|\ve x\right|$ and $\upsilon=\left|\ve\upsilon\right|$.
Now summing over all $N$ particles in the container of volume $V$
and assuming this volume is sufficiently large that it contains a
sufficiently large number of particles to realise a statistically
steady state, 
\begin{equation}
\frac{1}{4}\sum\left(\frac{d^{2}x^{2}(t)}{dt^{2}}+\beta\frac{dx^{2}(t)}{dt}\right)=\sum\left(\frac{1}{2}\upsilon^{2}+\frac{1}{2}\ve x(t)\cdot\ve f(t)\right)+\frac{1}{2}m^{-1}\sum\ve F_{e}\cdot\ve x(t).\label{eq:-1}
\end{equation}
 The value $x^{2}(t)$ averaged over all the particles will not change
with time at equilibrium since the particles are confined within the
walls of the containment and so the derivatives of the average value
$x^{2}$ will be zero%
\footnote{we are assuming that volume averages and derivatives commute %
}. So rearranging the RHS we can write this equation as 
\begin{equation}
\frac{1}{2}mV\left\langle n\right\rangle \left(\left\langle \upsilon^{2}\right\rangle +\left\langle \ve x(t)\cdot\ve f(t)\right\rangle \right)=-\frac{1}{2}\sum\ve F_{e}\cdot\ve x(t).\label{eq:The Virial Equation}
\end{equation}
 Eq.(\ref{eq:The Virial Equation}) is the Virial Equation and the
term on the RHS often referred to as the Virial, where $\left\langle n\right\rangle $
is the average number density in the container $N/V$ and $\left\langle \upsilon^{2}\right\rangle $is
the net kinetic energy per unit mas of particles $N^{-1}\sum\upsilon^{2}$.
The term on the RHS involves an integration over the total stress
at the walls of the container. In this case the stress is $-p$ in
the direction normal to the surface $S$ of the containment. So
\begin{equation}
\sum\ve F_{e}\cdot\ve x(t)=-p\int_{S}\ve x\cdot d\ve S=-p\int_{V}\ve\nabla\cdot\ve x\, dV=-3pV.\label{eq:-2}
\end{equation}
So Eq.(\ref{eq:The Virial Equation}) can be written as 
\begin{equation}
\frac{1}{2}mV\left\langle n\right\rangle \left(\left\langle \upsilon^{2}\right\rangle +\left\langle \ve x(t)\cdot\ve f(t)\right\rangle \right)=\frac{3}{2}pV\label{eq:-3}
\end{equation}
which finally gives the equation of state for the suspended particles
, namely
\begin{equation}
\frac{p}{\left\langle \rho\right\rangle }=\frac{1}{3}\left\langle \upsilon^{2}\right\rangle +\frac{1}{3}\left\langle \ve x(t)\cdot\ve f(t)\right\rangle \label{eq:Equation of State}
\end{equation}
where $\left\langle \rho\right\rangle $ i.e. the average mass density
of the suspended particles, $m\left\langle n\right\rangle $(see Eq.
(9) of Reeks\cite{Reeks91}). We note that from the solution of Eqs.(\ref{eq:particle equation of motion})
for $t\rightarrow\infty,$i.e. equilibrium conditions, so 
\begin{equation}
\left\langle \upsilon^{2}\right\rangle =\beta^{-1}\intop_{0}^{\infty}e^{-\beta s}\left\langle \ve f(0\cdot\ve f(s)\right\rangle ds\;;\;\left\langle \ve x(t)\cdot\ve f(t)\right\rangle =\beta^{-1}\intop_{0}^{\infty}(1-e^{-\beta s})\left\langle \ve f(0\cdot\ve f(s)\right\rangle ds\label{eq:-4}
\end{equation}
 and substituting in Eq.($\ref{eq:Equation of State}$) gives finally
\begin{equation}
\frac{p}{\left\langle \rho\right\rangle }=\frac{1}{3}\beta^{-1}\intop_{0}^{\infty}\left\langle \ve f(0\cdot\ve f(s)\right\rangle ds.\label{eq: Reduced form of eq.of State}
\end{equation}
\\
We note that in \cite{Reeks91} the quantity on the right hand side
of Eq.(\ref{eq: Reduced form of eq.of State}) was referred to as
the analogue of temperature, not the kinetic energy per unit mass
of the particles as it would be if we were dealing with molecules
in a gas. This would only be the case for very inert particles $St\gg1$,
when $\ve f(t)$ corresponds to a white noise driving force as is
the case for Brownian motion. \\
We recall also in \cite{Reeks91} the analogy that was drawn of the
equation of state for the suspended particles with that of a real
gas where the pressure is reduced from its ideal gas value by contributions
to the virial from the intermolecular forces. For the dispersed phase
the pressure, caused by the particle motion, is enhanced by contributions
to the virial from net accelerations induced by the fluctuating interphase
force (per unit volume), in this case $\left\langle \rho\ve f(\ve x,t\right\rangle $
where $\ve f(\ve x,t)$ is the driving force (per unit mass of particles)
experienced by particles in an elemental volume of the dispersed phase
mixture. In fact we can use the form of $p$ in Eq.(\ref{eq:Equation of State})
to evaluate this term as the dispersed phase approaches equilibrium.
The net momentum equation at equilibrium for an elemental volume of
the gas-particle mixture would be given by 
\begin{equation}
-\frac{\partial}{\partial x_{i}}\left\langle \rho\upsilon_{i}\upsilon_{j}\right\rangle +\left\langle \rho f_{j}\right\rangle =0\label{eq:-5}
\end{equation}
(see Eq.(10) in Reeks (1991) \cite{Reeks91}). In the case of the
suspended particles in an isotropic turbulent flow$\left\langle \rho\upsilon_{i}\upsilon_{j}\right\rangle =\frac{1}{3}\left\langle \rho\upsilon^{2}\right\rangle \delta_{ij}$
\footnote{$-\left\langle \rho\upsilon_{i}\upsilon_{j}\right\rangle $is often
referred to as the kinetic stresses equivalent to the Reynolds stresses
in turbulence modelling.$\frac{1}{3}\left\langle \rho\upsilon^{2}\right\rangle $
could similarly be referred to as the kinetic pressure %
} so 
\begin{equation}
-\frac{1}{3}\frac{\partial}{\partial x_{j}}\left\langle \rho\upsilon^{2}\right\rangle +\left\langle \rho f_{j}\right\rangle =0.\label{eq:HIT particle momentum equation}
\end{equation}
The equilibrium condition implies that the pressure defined in Eq.
(\ref{eq:Equation of State}) is uniform which means that 
\begin{equation}
-\frac{\partial}{\partial x_{j}}p=0\label{eq:-6}
\end{equation}
which substituting the expression for $p$ given in the equation of
state Eq.(\ref{eq:Equation of State}) means 
\begin{equation}
-\frac{1}{3}\frac{\partial}{\partial x_{j}}\left(\left\langle \rho\right\rangle \left\langle \upsilon^{2}\right\rangle +\left\langle \ve x(t)\cdot\ve f(t)\right\rangle \left\langle \rho\right\rangle \right)=0\label{eq:grad of pressure involing kinetic pressure}
\end{equation}
so for the force balance in Eq.(\ref{eq:HIT particle momentum equation})
to be equivalent to a uniform pressure at equilibrium expressed explicitly
in Eq.(\ref{eq:grad of pressure involing kinetic pressure}) , $\left\langle \rho f_{j}\right\rangle $must
also be equivalent the gradient of pressure (or in general in situations
where the flow is homogeneous but not isotropic to the gradient of
a stress tensor) can be interpreted as a diffusive flux for which
$\frac{1}{3}\left\langle \ve x(t)\cdot\ve f(t)\right\rangle $is the
diffusion coefficient. If $f_{i}=\beta u_{i}$ i.e. Stokes drag, then
\begin{equation}
\left\langle \rho u_{i}\right\rangle =-\frac{1}{3}\left\langle \ve x(t)\cdot\ve u(t)\right\rangle \frac{\partial}{\partial x_{j}}\left\langle \rho(\ve x,t\right\rangle \label{eq:-7}
\end{equation}
 and this case $\frac{1}{3}\left\langle \ve x(t)\cdot\ve u(t)\right\rangle $
is what has been referred to as the particle-fluid diffusion coefficient.
The density weighted flow velocity $\overline{\ve u}=\left\langle \rho u_{i}\right\rangle /\left\langle \rho\right\rangle $is
necessarily the net flow velocity sampled by a particles in an elemental
volume of the carrier flow $\ve x,t$\\
\\
Finally we recall here the way in \cite{Reeks91} the equation of
state for the suspended particles at equilibrium was used to evaluate
the long term particle diffusion using exactly the same method that
Einstein \cite{Einstein1905} used to evaluate the diffusion coefficient
of Brownian particles. Here we have an almost identical particle equation
of motion Eq.(\ref{eq:particle equation of motion}) except the driving
force (due to the turbulence carrier flow) is not limited to white
noise as it is in the case of Brownian motion due to molecular bombardment
of the suspended particles. What Einstein recognised was that the
momentum equation (in his case the balance of the pressure gradient
with the weight of the particles) implies a diffusion equation for
the suspended particles as they approached their long terms equilibrium
state and in particular as the average particle concentration $\ve\nabla\left\langle \rho\right\rangle \rightarrow0$.
So instead of an isothermal system, we have a statistical stationary
homogenous isotropic turbulent flow and we consider an equilibrium
state in which there is a balance between the pressure gradient and
a body force acting on the particles, the obvious one being the weight
of the particles, so in effect we are considering the weight of an
elemental volume of particles balanced by the pressure gradient acting
across it. So if $g$ is the acceleration due to gravity (force per
unit mass) acting in the $x_{i}$ direction, then this implies that
\begin{equation}
g\left\langle \rho\right\rangle -\frac{\partial p}{\partial x_{i}}=0\label{eq:-8}
\end{equation}
which substituting the expression for $p$ in Eq.(\ref{eq: Reduced form of eq.of State})
we have
\begin{equation}
g\left\langle \rho\right\rangle -\frac{1}{3}\beta^{-1}\intop_{0}^{\infty}\left\langle \ve f(0\cdot\ve f(s)\right\rangle ds\frac{\partial\left\langle \rho\right\rangle }{\partial x_{i}}=0.\label{eq:withe paressure gradient balance}
\end{equation}
Alternatively we could consider as Einstein did for Brownian motion,
this equilibrium as a balance between a convection current $\beta^{-1}g\left\langle \rho\right\rangle $and
a diffusion current $-\epsilon(\infty)\frac{\partial\left\langle \rho\right\rangle }{\partial x_{i}}$
where $\epsilon(\infty)$ denotes the long term particle diffusion
coefficient. Thus 
\begin{equation}
\beta^{-1}g\left\langle \rho\right\rangle -\epsilon(\infty)\frac{\partial\left\langle \rho\right\rangle }{\partial x_{i}}=0.\label{eq:balance of drft + diffusion}
\end{equation}
So assuming Eqs.(\ref{eq:balance of drft + diffusion}) is the same
as Eq. (\ref{eq:withe paressure gradient balance}) we must have 
\begin{equation}
\epsilon(\infty)=\frac{1}{3}\beta^{-2}\intop_{0}^{\infty}\left\langle \ve f(0\cdot\ve f(s)\right\rangle ds.\label{eq:long time particle diffusion coefficient}
\end{equation}
 The interesting result is the case of Stokes drag in which case $\ve f=\beta\ve u$
and 

\begin{equation}
\epsilon(\infty)=\frac{1}{3}\intop_{0}^{\infty}\left\langle \ve u(0)\cdot\ve u(s)\right\rangle ds,\label{eq:Stokes  drag diffusion coefficient}
\end{equation}
indicating no explicit dependence on particle inertia a result derived
by more formal means using Taylor's formula for the particle diffusion
coefficient, namely 
\begin{equation}
\epsilon(\infty)=\frac{1}{3}\intop_{0}^{\infty}\left\langle \ve\upsilon(0)\cdot\ve\upsilon(s)\right\rangle ds,\label{eq:Taylor's formula}
\end{equation}
and then substituting the integral expression for the particle velocity
$\ve\upsilon$ involving $\ve u(s)$ from $s=0,t$ , giving the surprising
result in Eq.(\ref{eq:Stokes  drag diffusion coefficient}) (confirmed
by DNS of particle dispersion in an isotropic turbulent flow\cite{Squires&Eaton91}.
This lack of inertia dependence is in contrast to that for the Brownian
diffusion coefficient $\epsilon_{B}$ which from the Stokes-Einstein
relation is 
\begin{equation}
\epsilon_{B}=k_{B}T/m\beta.\label{eq:-10}
\end{equation}
\\
We can also obtain the same result for the particle diffusion coefficient
without invoking the addition of an extra body force, by considering
the long term dispersion of particles into an infinite flow (no boundaries).

\footnote{Note this is different from considering the suspension of particles
at equilibrium within some confined space although it comes to the
same result in the end. However it does define the timescales for
which the suspended particles approach equilibrium (rather than arbitrarily
saying $t\rightarrow\infty)$. In the equilibrium case we began with,
the particles are contained within a finite volume by the walls of
the containment which exert a pressure on the particles to maintain
that confinement. In the long term dispersion case there are no boundary
conditions imposed but as time $t\rightarrow\infty$ the particles
approach an equilibrium condition within a finite volume of the particles
but necessarily one in which although the concentration is reducing,
the concentration within the volume approaches a uniform value. We
could call this quasi-equilibrium. In this case the mean velocity
of the particles approaches zero and the mean drag is balanced by
the pressure gradient. This requires from Eq.(\ref{eq:momentum equaition})
that in general $\beta^{-1}\overline{\upsilon}_{i}^{-1}\mbox{\ensuremath{D}\ensuremath{\overline{\upsilon}_{i}}/\ensuremath{Dt}\textasciitilde\ensuremath{\beta^{-1}\epsilon}/\ensuremath{L}}^{2}.$
$L^{2}\sim\epsilon t$, which implies that $\beta t\gg1$.%
} The momentum equation can be written as 
\begin{equation}
\left\langle \rho\right\rangle \frac{D\overline{\upsilon}_{i}}{Dt}=-\frac{\partial p}{\partial x_{i}}-\beta\overline{\upsilon}_{i}\left\langle \rho\right\rangle .\label{eq:momentum equaition}
\end{equation}
Recognizing that $\overline{\upsilon}_{i}\left\langle \rho\right\rangle $
is the diffusion flux, we can write Eq.(\ref{eq:momentum equaition})
as 
\begin{equation}
\overline{\upsilon}_{i}\left\langle \rho\right\rangle =\beta^{-1}\frac{\partial p}{\partial x_{i}}-\beta^{-1}\left\langle \rho\right\rangle \frac{D\overline{\upsilon}_{i}}{Dt}.\label{eq:-12}
\end{equation}
Assuming that the inertial acceleration terms on the RHS can be ignored
compared to the other terms and that in the long term limit (satisfied
if $\beta t\gg1)$ then we have a balance between the drag force acting
on an elemental volume of particles and the pressure gradient. Replacing
$p$ with the expression given in Eq. (\ref{eq: Reduced form of eq.of State})
gives the value for the long time particle diffusion coefficient$\epsilon(\infty)$
given in Eq.(\ref{eq:long time particle diffusion coefficient}) for
which in the long time $\beta t\rightarrow\infty$, we obtain Fick's
Law for particle diffusion
\begin{equation}
\overline{\upsilon}_{i}\left\langle \rho\right\rangle \rightarrow-\epsilon(\infty)\frac{\partial\left\langle \rho\right\rangle }{\partial x_{i}}.\label{eq:-13}
\end{equation}
Note there is a self consistency here, in that $\beta t\gg1$ means
particles have lost all memory of their initial conditions, and when
$\epsilon(\infty)$ is formally derived from the equation of motion
and using of Taylor's formula Eq.(\ref{eq:Taylor's formula}, a similar
condition applies. Of course there is also the implicit assumption
that $t/\tau_{f}\gg1$.

\end{document}